\documentclass[12pt]{article}
 \usepackage{epsfig}
 \newcommand{\be}{\begin{equation}}
 \newcommand{\bea}{\begin{eqnarray}}
 \newcommand{\eea}{\end{eqnarray}}
 \newcommand{\ba}{\begin{array}}
 \newcommand{\ea}{\end{array}}
 \newcommand{\ee}{\end{equation}}

 \expandafter\ifx\csname mathbbm\endcsname\relax

 \else

 \fi
\begin{document}
 \begin{titlepage}
 \hfill
 \vbox{
    \halign{#\hfil         \cr
           hep-th/0408015 \cr
           IPM/P-2004/037 \cr
           } 
      }  
 \vspace*{20mm}
 \begin{center}
 {\Large {\bf D-brane Dynamics in RR Deformation 
 of NS5-branes Background and Tachyon Cosmology}\\ }

 \vspace*{15mm} \vspace*{1mm} {Ahmad Ghodsi and Amir E. Mosaffa}

 \vspace*{1cm}

 {\it  Institute for Studies in Theoretical 
 Physics and Mathematics (IPM)\\
 P.O. Box 19395-5531, Tehran, Iran\\}

 \vspace*{1cm}
 \end{center}

 \begin{abstract}
 We study D-brane dynamics in the gravity dual
 background of ODp theories using the effective
 action on the worldvolume of the brane. We 
 explore the similarities of the system with 
 the rolling tachyon, including the exponential 
 decrease of pressure at late times. We also 
 consider the case where the worldvolume theory on a
 D$_3$-brane is coupled to gravity and construct 
 a cosmological model. By considering the time 
 reversal symmetric solutions, we find that we can 
 have both closed and open universes depending on
 the initial value of the radial mode on the brane. 
 We compare the models with tachyon driven 
 cosmologies and find limits where
 slow roll inflation is realized. 
 \end{abstract}

 \end{titlepage}

 \section{Introduction}
 D-brane interactions has been a subject of interest 
 for a long time. The subject has received more attention 
 since it was suggested that the worldvolume theory on 
 D-branes can, in some cases, be used to construct 
 inflationary models for cosmology (see  e.g 
 \cite{Verlinde:1999fy},\cite{Kehagias:1999vr},\cite
 {Kachru:2003aw},\cite{Kachru:2003sx}). 
 To construct such models one requires the D-brane 
 to go through a time dependent process in string 
 theory. In some cases, this time dependence is 
 provided by considering branes with tachyonic modes 
 which cause an instability in the system.
 The unstable brane decays as the tachyon, which is 
 playing the role of the inflaton, rolls down a
 potential. Such models are known as tachyon driven 
 cosmologies (see \cite{Sen},\cite{tachcosmo},\cite{alex}). 
 In other cases, the time dependence is achieved by 
 considering non BPS configurations where the D-brane 
 moves under the influence of a potential. The inflaton 
 is then identified with the transverse mode that specifies 
 the location of the brane. 
 A considerable amount of work has been done recently using 
 this scenario \cite{Kachru:2003sx} and its 
 modifications \cite{warpcosmo},\cite{eva},\cite{aa}.
 
 As an example of such a non BPS
 configuration, one can place a D-brane in the
 background produced by NS5 branes 
 in such a way that no supersymmetry is preserved.
 NS5 branes are special in a number of ways.
 An important fact about these objects is 
 that unlike RR charged D-branes, they
 keep curving the space around them even 
 when $g_s\rightarrow0$. This can also be seen
 by the fact that the tension of the D-brane 
 goes as $1/g_s$, whereas
 for the fivebranes it goes as $1/g_s^2$ so 
 that the latter becomes much heavier
 than the former in this limit. Therefore NS 
 fivebranes are known to be solitonic
 rather than perturbative objects in 
 string theory. Another important issue
 is that the string coupling becomes exceedingly 
 large as one approaches the bottom
 of the throat produced by these objects.

 The dynamics of a D-brane, parallel
 to a stack of NS5-branes was studied in 
 \cite{Kutasov:2004dj} using the effective action 
 on the worldvolume of the D-brane, 
 the DBI action\footnote{see also \cite{ns5} 
 for earlier works and \cite{Panigrahi:2004qr}
 for the D$_p$-brane background.}.
 Based on the above argument,
 the D-brane can be considered as a probe that 
 is exploring the background without
 influencing it considerably.
 It was shown that for a certain range of 
 energies of the D-brane, most
 of the dynamical process occurs when the 
 coupling is still small
 and perturbative string theory is reliable.
 Since this configuration of branes breaks all
 of the supersymmetry, the probe experiences 
 a net force from the background
 and becomes unstable. This is reminiscent 
 of the rolling tachyon problem. It was shown in
 \cite{Kutasov:2004dj} that by a proper field 
 redefinition, the effective action of the probe 
 can actually be brought into the form of the 
 DBI action for the tachyon in the open string
 models \cite{garousi},\cite{effective}. Being 
 related to the transverse coordinate of the brane, 
 the defined tachyon field finds a geometric 
 interpretation in this case.
 By looking at the energy momentum tensor of the 
 worldvolume theory, it was shown that as the 
 brane approaches the fivebranes the pressure 
 goes exponentially to zero which is the late 
 time behavior of tachyon matter \cite{Sen:2002an}. 
 For the tachyon, this is interpreted as the 
 decay of the unstable brane into closed string 
 radiation \cite{closedradiation}. For the probe 
 D-brane, it was argued that the pressureless fluid 
 is a signal that the brane sheds its energy into modes 
 living on the fivebranes. Losing almost all of its 
 energy as it approaches the fivebranes, the
 D-brane will eventually form a bound 
 state with the fivebranes.

 In the present work we study the effect of RR fields in
 the above processes. For this purpose, we consider the
 deformation of NS5-brane background in presence 
 of nonzero RR forms \cite{RR1},\cite{RR2}. We 
 focus on the limit where the theory on the fivebranes
 decouples from the bulk. In the absence of these 
 RR fields, the resulting theory is known as Little 
 String Theory (LST) \cite{LST}. The RR fields deform 
 LST to what is known as ODp (Open D$_p$-brane) 
 theories. We will study the dynamics of 
 a D-brane in the background which is the gravity 
 dual of ODp theories.

 RR fields change things in two different ways. 
 Firstly, the metric and dilaton background fields 
 are deformed and secondly, a term is added to the 
 effective action describing the coupling of
 the D-brane to the RR background fields. It will 
 turn out that the behavior of the D-brane, including 
 its similarities with the rolling tachyon
 problem, remains unchanged at late times.

 We will also consider the case where the effective 
 action on the worldvolume of a probe D$_3$-brane, which 
 is moving in the NS5 branes background, is coupled 
 to gravity \footnote{for a related recent work see 
 \cite{Yavartanoo:2004wb}.}by introducing a dynamical
 four dimensional metric and thus making a cosmological 
 model. For the metric, we choose it to be the 
 FRW solution of Einstein's equations.
 We also include a small positive cosmological 
 constant term in the action to allow for nonsingular 
 solutions. We find that if we place the brane far 
 from the fivebranes the resulting solution will be 
 an open universe. For small enough distances as
 well, such an open solution is obtained. 
 But there is an intermediate range of distances 
 from the fivebranes where, if the D-brane starts
 from, a singularity will occur in the solution 
 and thus the universe will close.

 The deformation of the background by RR fields 
 modifies the cosmological solutions slightly. 
 Although in the UV regime the potential for 
 rolling is affected by the deformation, the 
 general behavior of the solutions remains unchanged. 
 By analyzing the equations of motion, we comment 
 on the relative width of the initial values that 
 the tachyon field can take in order to result 
 in singular solutions. We support our judgments by
 presenting some of the values which are obtained by 
 numerical analysis. In each case we will compare 
 our solutions with the ones obtained for rolling 
 tachyon and observe that although at late times 
 the similarities of the two systems persist in 
 the coupled to gravity theories, we can find limits 
 where unlike the rolling tachyon, the criteria 
 for slow roll inflation are met.

 \section{\large{D-branes in the NS5-branes Background in 
 presence of RR fields}}
 In this section we study the dynamics of a D-brane in the
 noncommutative deformation of Little String Theory (LST).
 Let us first remind some basic facts about type II 
 NS5-branes and its decoupling limit. The solution 
 describing $N$ coincident NS5-branes is \cite{Callan:1991at}
 \bea
 ds^2&=&dx_\mu
 dx^\mu+h(x^n)dx_m\,dx^m\;,\cr &&\cr
 e^{2\Phi-2\Phi_0}&=&h(x^n)\;,\cr &&\cr
 H_{mnp}&=&-\epsilon^q_{mnp}\partial_q\Phi\;.
 \label{bac}
 \eea
 where $x^\mu$ are tangent to the fivebranes and
 $x^m$ are transverse to them. $H_{mnp}$ is the field
 strength of the NSNS two form and the harmonic function 
 $h$ is given by
 \be 
 h=1+\frac{N\,l_s^2}{r^2}\;, \label{f}
 \ee
 where $r=\sqrt{x_n\,x^n}$. Being much
 lighter than the solitonic fivebranes in 
 the small string coupling
 regime, the D-brane can be considered as a 
 probe exploring the above background.
 The metric describes a geometry which
 is flat along the fivebranes and is 
 isotropic in the transverse
 directions with an $SO(4)$ symmetry. There is 
 an infinite throat with the geometry $S^3\times 
 R^1\times$ a six dimensional Minkowski space. 
 As one moves down the throat towards the
 fivebranes at $r=0$, the radius of the $S^3$ 
 approaches an asymptotic value which is proportional 
 to $\sqrt{N}\;l_s$. This is the characteristic 
 length or the radius of the solution. Due to
 nonrenormalization theorems, the solution (\ref{bac}) 
 remains a valid string background even if its 
 scale is the string scale i.e if $N$ is small. 
 The dilaton grows in the throat, diverging as
 one approaches the fivebranes and hence a probe 
 D-brane traveling down the throat will sooner or 
 later move in the large dilaton region
 where string perturbation theory breaks down. 
 Yet, as was mentioned in \cite{Kutasov:2004dj}, 
 by assuming certain energies for the brane one 
 can postpone this to very late times.

 The decoupling limit of (\ref{bac}), known as the 
 linear dilaton background, is obtained by taking the 
 limit $g_s\rightarrow0$ and keeping $l_s$ and 
 $u\equiv{\frac{r}{g_s}}$ fixed where (\ref{bac})
 reduces to
 \bea
 ds^2&=& dx_\mu dx^\mu+{\frac{Nl_s^2}{u^2}}
 (du^2+u^2d\Omega_3^2)\;,\cr &&\cr
 e^{2\phi}&=&{\frac{Nl_s^2}{u^2}}\;,\;\;\;\;\;
 dB=2N\epsilon_3\;.
 \label{lindil}
 \eea
 This background provides the gravity dual 
 for Little String Theory (LST) which lives 
 on the worldvolume of the fivebranes
 when decoupled from the bulk \cite{LST}.

 One can deform LST by adding RR fields to the story. 
 What happens in practice is that one first obtains 
 the solution describing NS5-branes in presence of 
 nonzero RR fields and then takes the decoupling 
 limit of the resulting background.
 It turns out that in order to include an RR
 electric $p+1$-form in the problem one has to 
 add an RR magnetic $5-p$-form as well and vice 
 versa ($p=0,\cdots,5$). Hence the theory with 
 electric $p+1$-form is the same as that with 
 magnetic $5-p$-form.
 The decoupling limit, though, is quite
 different from what we saw above (which 
 resulted in the linear dilaton background). 
 The resulting background provides the
 gravity dual description for ODp theories 
 which live on the worldvolume of NS5-branes 
 and the excitations of which include
 light open Dp-branes \cite{odp},\cite{RR2}.

 After doing all this, the solution describing $N$ 
 coincident NS5-branes in presence of nonzero 
 RR fields after taking the decoupling limit 
 is \cite{RR2}
 \bea
 ds^2&=&(1+a^2\,r^2)^{1/2}\left[-dt^2+\sum_{i=1}^{p}
 dx_{i}^2+\frac{\sum_{k=p+1}^{5}d x_{k}^2}{1+a^2\,r^2}
 +\frac{N}{r^2}\,dx_m\,dx^m\right]\ ,\cr &&\cr
 A_{0\cdots p}&=&{\frac{1}{{\tilde g}}} a^2\,r^2\ \ ,
 \;\;\;\;\;\;\;\;\;\; \;\;\;\;\;\; A_{(p+1)\cdots 5}
 ={\frac{1}{{\tilde g}}}\;\frac{a^2\,r^2}{1+a^2\,r^2}\
  , \cr &&\cr
 e^{2\phi}&=&{\tilde g}^2 \frac{(1+a^2\,r^2)^
 {(p-1)/2}}{a^2\,r^2},\;\;\;\;\;\;\;dB=2N\;
 \epsilon_3,\;\;\;\; a^2={\frac{l_{\rm eff}^2}{N}}\;,
 \label{nbc}
 \eea
 where $p=1,3,5$ in IIB and $p=0,2,4$ in IIA theory,
 $r=\sqrt{x^mx^m}$ and $l_{eff}$ and $\tilde g$ 
 are effective string tension and coupling after 
 the decoupling limit respectively. In the above 
 solution we have dropped powers of $l_s$ that 
 appear as an overall factor in the metric and in the
 definitions for the RR fields because they 
 will play no role in the equations of motion 
 for the probe branes.

 \subsection{Setup of the problem}
 We will now study the dynamics of a 
 probe D$_{p'}$-brane in the
 background (\ref{nbc}), using the effective 
 action on the worldvolume of the D-brane. 
 The embedding of the D-brane in the bulk space is
 expressed through the maps 
 $X^A(\xi^\mu)$ ($A=0,1,...,9$). We choose the
 coordinates of the probe, $\xi^\mu$ 
 $(\mu=0,1,...,p')$, to lie completely inside 
 the directions of the original NS5-brane so that 
 it is pointlike in the directions transverse to 
 them, $X^m(m=6,7,8,9)$, which we parameterize as
 \be
 \sum_{m=6}^{9} (dX^{m})^2=dr^2+ r^2 (d \phi^2+
 \sin^2\phi\; d \theta^2+ \cos^2\phi \;d \psi^2)\;.
 \ee
 We also assume that $X^m$'s depend only on the 
 time coordinate of the worldvolume,
 $\xi^0$, such that $X^m(\xi^\mu)=X^m(\xi^0)$. 
 The other transverse coordinates of the 
 D-brane (if any), we assume to be constant. 
 One can also in general have gauge fields on the
 D-brane which, in this work, we always assume 
 to be zero. Therefore the only dynamical fields 
 of the worldvolume effective theory will be $X^m(\xi^0)$.

 The effective action of a D-brane consists 
 of a DBI part and CS terms
 arising from coupling to the RR fields
 \be
 S_p=-\tau_p\int d^{p+1}\,\xi\, e^{-(\Phi-\Phi_0)}
 \sqrt{-det(G_{\mu\nu}+B_{\mu\nu})}+\tau_p\int 
 d^{p+1}\xi\, A \;,
 \label{dbi}
 \ee
 where $G_{\mu\nu}$ and $B_{\mu\nu}$ are the 
 metric and the NS two form induced
 on the worldvolume respectively and $A$ is 
 potential form for RR fields. For the ansatz we 
 have taken $B$ will always be zero.

 In the cases we study in the following sections, 
 the action (\ref{dbi}) will eventually lead to 
 a Lagrangian, $\cal {L}$, describing the motion of a
 particle moving in a central force field. Thus the
 motion will totally take place in a 2-plane 
 which, without loss of generality, we take to 
 be at $\phi={\frac{\pi}{2}}$ and hence we are 
 left with only two dynamical fields, $r$ 
 and $\theta$. There are two conserved quantities 
 associated with the motion of the D-brane;
 energy ($E$) and angular momentum ($L$)
 \be
 E=\frac{\partial \cal{L}}{\partial 
 \dot{X}^m} \;\dot{X}^m- {\cal{L}}\;,
 \;\;\;\;\;\;\; L=\frac{\partial \cal{L}}{\partial
 \dot{\theta}}\;,
 \ee
 where dot means differentiation with respect 
 to $\xi^0$. The above relations enable us to 
 find $\dot{r}$ in terms of $r$, $E$ and $L$ 
 and to define an effective potential,
 $V_{eff}(r)$, which describes a particle of 
 mass $=2$ and zero energy moving in
 one dimension. Cases with $L=0$ describe 
 purely radial motion.

 There are three distinct cases that we 
 consider. In the first two cases the 
 $\xi^\mu$'s lie inside the $(t,x^i)$
 \footnote{$1\leq i\leq p $ and $p+1\leq k\leq 5$} 
 part of the spacetime; in the first case 
 $p'<p$ whereas in the second one $p'=p$. 
 In the third case the $\xi^\mu$'s fill some or 
 all of the directions $(t,x^i,x^k)$.

 The reliability of this analysis requires 
 small proper acceleration for the scalar
 fields on the worldvolume of the
 D-brane and a small string coupling at its location.

 \subsection{Case 1}
 In this case the $\xi^\mu$ fill some but not all of the
 directions $(t,x^i)$ such that $p'< p$. We 
 should necessarily have
 $2\leq p\leq5$. The reparametrization
 invariance of the worldvolume theory 
 allows us to choose $\xi^\mu=x^\mu$.
 There will be no coupling to the RR fields 
 in this case and the effective action
 will read
 \be
 S_{p'}=-a\,\tau_{p'}\,\int
 d^{p'+1}\,\xi
 \;f(r)^{\frac{j}{4}}\;r\;\sqrt{1-
 \frac{N}{r^2}\dot{X}^m\,\dot{X}^m}\;,
 \label{nac}
 \ee
 where $f(r)=1+a^2\,r^2$, $j=p'-p+2$ and dot 
 now stands for $t$ derivatives.

 Energy and angular momentum densities will read
 \bea
 E&=&\frac{a\,\tau_{p'}\,r\,f^{\frac{j}{4}}}
 {\sqrt{1-\frac{N}{r^2}(\dot{r}^2+r^2\,\dot{\theta}^2)}}\;,\cr
 &&\cr L&=&\frac{a\,\tau_{p'}\,N\,r\,\dot{\theta}
 f^{\frac{j}{4}}}{\sqrt{1-\frac{N}{r^2}
 (\dot{r}^2+r^2\,\dot{\theta}^2)}}\;.
 \label{nel}
 \eea
 Defining $u\equiv a\;r$ we have
 \bea
 \dot{u}^2&=&\frac{u^2}{N}\bigg(1-\frac{L^2}{E^2\,N}
 -\frac{u^2}{\lambda_{p'}^2}(1+u^2)^{\frac{j}{2}}\bigg)
 \;,\cr &&\cr
 e^\Phi&=&\frac{\tilde{g}}{u}(u^2+1)^{\frac{p-1}{4}}\;,
 \label{emsc}
 \eea
 where $\lambda_{p'}=\frac{E}{\tau_{p'}}$.
 One should note that for both IIA and 
 IIB theories $j=-2,0$. As the D-brane goes from 
 large values of $u$ towards $u=0$, the
 coupling increases for $p=2,3$ monotonically 
 whereas for $p=4,5$ it decreases until 
 $u=u_g\equiv\sqrt{\frac{2}{p-3}}$
 and then increases as it gets closer to $u=0$.

 The effective potential is now defined 
 as $V_{eff}(u)=-\dot{u}^2$. Qualitative
 description of motion can be achieved by 
 finding the general shape of
 the potential, for which, we look at its 
 behavior in small and large
 values of $u$ for each value of $j$. 
 
 $ $

 {\bf(A):}$\,\,j=-2\,,\,L\neq0$
 \bea
 V_{eff}&\rightarrow&\frac{u^2}{N}
 \bigg(\frac{L^2}{E^2\,N}-1\bigg)
 \;\;\;\;\;\;\;\;\;\;\;\;\;\;\;
 \;\;\;\;\;\; u\rightarrow 0\;\;,\cr &&\cr
 V_{eff}&\rightarrow&\frac{u^2}{N}
 \bigg(\frac{1}{\lambda_{p'}^2}
 +\frac{L^2}{E^2\,N}-1\bigg)
 \;\;\;\;\;\;\;\;\;\;\;\;u\rightarrow \infty\;.
 \label{vjzl}
 \eea
 There are three regimes for energy:
 
 $ $

 {(A.1):} $\,\,\,\tau_{p'}^2+\frac{L^2}{N}<E^2$;
 the D-brane explores all values of u.
 It can either start
 at $u=0$ at early times and escape to infinity at 
 late times or vice versa.
 
 $ $
 
 {(A.2):} $\,\,\,\frac{L^2}{N}<E^2<\tau_{p'}^2
 +\frac{L^2}{N}$; the brane can get at most as 
 far as $u_{max}=\lambda'_{p'}\equiv \lambda_{p'}
 \sqrt{1-\frac{L^2}{E^2}}$ from $u=0$. If the brane
 starts at  $u=u_{max}$ at $t=0$, it will take 
 it an infinite amount of time to fall towards the origin.
 If $\lambda_{p'}\ll 1$, the brane always remains 
 at small values of $u$. So one can use the
 approximation of $V_{eff}$ in this region to solve 
 for $u$ with the result
 \be 
 \frac{1}{u}=\frac{1}{\lambda'_{p'}}
 \cosh\;\bigg(\sqrt{1-\frac{L^2}{E^2}}\;\frac{t}
 {\sqrt{N}}\bigg)\;.
 \label{soljzl}
 \ee
 The string coupling in this region is approximated as
 \be 
 e^\Phi\approx\frac{\tilde{g}}{u}\;,
 \ee
 therefore if
 \be 
 \tilde{g}\ll\lambda_{p'}\ll1\;, \label{dil}
 \ee
 the solution (\ref{soljzl}) is 
 reliable for a considerably long period of time.
 By a calculation quite similar to 
 what is done in \cite{Kutasov:2004dj},
 one can see that the brane is in a 
 spiraling motion as it falls towards the origin.

 $ $
 
 {(A.3):} $\,\,\,E^2<\frac{L^2}{N}$; the 
 D-brane remains at $r=0$ for all times.

 $ $
 
 {\bf(B):}$\,\,j=-2\,,\,L=0.$
 $ $
 
 In terms of $\lambda_{p'}$, there are two cases:

 $ $
 
 {(B.1):} $\,\,\lambda_{p'}>1$; the case is 
 just like (A.1). The solution describes either an 
 incoming brane towards $u=0$ or an outgoing one
 escaping to infinity.
 
 $ $

 {(B.2):} $\,\,\lambda_{p'}<1$; the 
 situation is qualitatively that of case (A.2).
 Here the D-brane can get no further than
 $u_{max}=\frac{\lambda_{p'}}{\sqrt{1-\lambda_{p'}}}$ 
 from the origin. If $\lambda_{p'}\ll 1$, this 
 maximum distance is still in the small $u$
 region and the following solution will 
 be a good approximation
 \be
 \frac{1}{u}=\frac{1}{\lambda_{p'}}\cosh(\frac{t}
 {\sqrt{N}})\;,
 \label{soljz}
 \ee
 where we have assumed that the brane is at rest 
 at $t=0$. Here again
 we require that the string coupling
 remains small. So the solution (\ref{soljz}) 
 is reliable for a long period
 of time if $\tilde{g}\ll\lambda_{p'}\ll1$.

 $ $
 
 {\bf(C):} $\,\,j=0\,,\,L\neq0$

 \bea
 V_{eff}&\rightarrow&\frac{u^2}{N}\;
 \bigg(\frac{L^2}{E^2\,N}-1\bigg)\;\;\;\;\;\;\;\;
 \;\;\;\;\;\;u\rightarrow0\;\;,\cr
 V_{eff}&\rightarrow&\frac{u^4}{\lambda_{p'}^2\,N}
 \;\;\;\;\;\;\;\;\;\;\;\;\;\;
 \;\;\;\;\;\;\;\;\;\;\;\;\;\;\;\;\;
 u\rightarrow\infty\;.
 \label{vjl}
 \eea
 There are two cases to consider

 $ $
 
 {(C.1):} $\,\,E^2\leq\frac{L^2}{N}$; the 
 brane remains at $u=0$ at all times.

 $ $
 
 {(C.2):} $\,\,\frac{L^2}{N}\leq E^2$; the 
 brane has a turning point
 at $u_{max}=\lambda'_{p'}$.
 The solution is the same as (\ref{soljzl}) but 
 this time it is valid for all
 values of $\lambda_{p'}$. To have a small coupling 
 for a long time period we must
 have $\tilde{g}\ll\lambda_{p'}\ll1$.
 
 $ $
 
 {\bf(D):} $\,\,j=0\,,\,L=0.$

 There is only one solution here describing a
 brane moving in $0\leq u\leq u_{max}=\lambda_{p'}$.
 The solution is (\ref{soljzl}) with $L=0$ which 
 is valid for all values
 of $\lambda_{p'}$. Here again we assume 
 that $\tilde{g}\ll\lambda_{p'}\ll1$.

 \subsection{Case 2}
 In this case the $\xi^\mu$ fill 
 all of the directions $(t,x^i)$ and
 $p'=p$. The worldvolume theory is still 
 reparametrization invariant and one can 
 set $\xi^\mu=x^\mu$. This time there is a
 coupling to one of the RR fields and the 
 effective action reads $(f=1+a^2r^2)$
 \be
 S_p=-a\,\tau_p\,V_p\int dt\,\bigg(r\,f^{\frac 12}
 \sqrt{1-\frac{N}{r^2}\dot{X}^m\dot{X}^m}-ar^2
 \bigg)\;.
 \label{dbicss}
 \ee
 The conserved quantities are
 \bea E&=&\frac{a\,\tau_p\,N\,f^{\frac12}\,r\,
 \dot{\theta}}{\sqrt{1-\frac{N}{r^2}(\dot{r}^2+r^2\,
 \dot{\theta}^2)}}\;,\cr
 &&\cr L&=&\frac{a\,\tau_p\,r\,f^{\frac12}}
 {\sqrt{1-\frac{N}{r^2}(\dot{r}^2+r^2\,\dot{\theta}^2)
 }}-a^2\tau_p\,r^2\;.
 \label{elpp}
 \eea
 Then, in terms of $u$, the equation of motion becomes
 \be
 \dot{u}^2=\frac{u^2}{N}\bigg(1-\frac{L^2}
 {E^2(1+\frac{u^2}{\lambda_p^2})^2\,N}-
 \frac{u^2\,f}{\lambda_p^2(1+\frac{u^2}
 {\lambda_p^2})^2}\bigg)\;.
 \label{eqmpp}
 \ee
 Now again we find the general behavior of the
 effective potential by looking at its form for 
 small and large values of $u$. There are two 
 distinct cases to consider, $L\neq0$ and $L=0$.
 
 $ $
 
 {\bf(E):} $\,\,L\neq0$
 \bea
 V_{eff}&\rightarrow&\frac{u^2}{N}\bigg(\frac{L^2}
 {N\,E^2}-1\bigg) \;\;\;\;\;\;\;\;\;\;\;\;\;
 \;\;\;\;\;u\rightarrow0\;\;,\cr &&\cr
 V_{eff}&\rightarrow&\frac{1}{N}\bigg(1-2\,
 \lambda_p\bigg) \;\;\;\;\;\;\;\;\;\;\;\;\;
 \;\;\;\;\;\;\;\;u\rightarrow\infty\;.
 \label{ppv}
 \eea
 There are four regimes for energy:

 $ $
 
 {(E.1):} $\,\,\frac{\tau_p}{2}<E<\frac{L}
 {\sqrt{N}}$; if the brane starts at $u=0$ it 
 remains there at all times, otherwise, it
 can get no closer to the origin than
 \be
 u_{m}^2=\lambda_p^2\;\frac{(1-\frac{L^2}{E^2\,N})}
 {(1-2\lambda_p)}\;.
 \label{ppu0}
 \ee
 The brane is then deflected and escapes back 
 to infinity.

 $ $
 
 {(E.2):} $\,\,\lambda_p<1$ and $E<\frac{L}
 {\sqrt{N}}$; the brane remains at $r=0$ at all times.

 $ $

 {(E.3):} $\,\,\lambda_p>1$ and $\frac{L}
 {\sqrt{N}}<E$; the brane goes from $r=0$ 
 to infinity or vice versa.

 $ $
 
 {(E.4):} $\,\,\frac{L}{\sqrt{N}}<E<\frac
 {\tau_p}{2}$; the brane moves between $u=0$ 
 and a maximum, $u_{m}$, which is given by 
 (\ref{ppu0}). If  $\lambda_p\ll 1$ 
 then $u_{m}\ll 1$ and the approximate 
 solution for $u$ and the condition for 
 small coupling for a long period of time are 
 respectively (\ref{soljzl}) and (\ref{dil}) 
 with $p'=p$.

 $ $

 {\bf(F):} $\,\,L=0.$

 There are two cases:

 $ $
 
 {(F.1):} $\,\,\lambda_p<\frac12$; the brane 
 moves between $u=0$and a maximum $u_{max}=
 \frac{\lambda_p}{\sqrt{1-2\,\lambda_p}}$. If 
 $\lambda_p\ll1$ then the approximate solution 
 for $u$ and small coupling condition are
 found to be respectively (\ref{soljzl}) 
 and (\ref{dil}) with $L=0$ and $p'=p$.

 $ $
 
 {(F.2):} $\,\,\frac12<\lambda_p$; the brane 
 goes from $u=0$ to infinity or vice versa.

 \subsection{Case 3}
 In this case the brane is extended 
 in some or all of the $(t,x^i,x^k)$ directions.
 The worldvolume theory no longer has 
 reparameterization invariance
 and hence the D-brane dynamics does depend on 
 the way it is embedded in the bulk. Here 
 we consider the simple ansatz for the
 embedding such that $X^A$ with $0\leq A\leq p$ 
 are independent from those with $p<A\leq5$. 
 The brane thus fills $s+1$ of the $(t,x^i)$
 and $q-p$ of the $x^k$ directions 
 with $s+q-p=p'$. One can then use the
 reparameterization invariance in the two 
 subspaces to set $x^\mu=\xi^\mu\;(\mu=0\;,
 \cdots,s,p+1,\cdots,q)$.

 The effective action consists only of the 
 $DBI$ part
 \be
 S_{p'}=-a\tau_{p'}\int d^{p'+1}\xi \;r\,
 f(r)^{\frac j4}
 \sqrt{1-\frac{N}{r^2}\dot{X}^m\dot{X}^m}\;,
 \ee
 where $j=s-q+2$. It is interesting to note 
 that $j$  can only take the values $0,-2$.
 Hence we are back to the action (\ref{nac}) 
 and each solution in this case corresponds
 to one of the solutions found in case 2.

 \subsection{Comments on the solutions}
 The above results show that the D-brane 
 can have four possible
 qualitative behaviors. The 
 first one is staying at the origin.
 The second one is moving between the origin 
 and a maximum distance. The third one 
 is moving between the origin and infinity and 
 finally the forth possibility is to move
 between a minimum distance from $u=0$ 
 and infinity. The first and fourth of 
 these can only happen for a  brane with 
 a nonzero angular momentum. As was found 
 in \cite{Kutasov:2004dj}, all  four 
 possibilities can be experienced by 
 a D-brane in the background (\ref{bac}). 
 As a result of deformation, one finds 
 some cases that even when the brane 
 has a nonzero angular momentum, not 
 all of the four behaviors are observed. 
 For example, the fourth possibility is 
 excluded in case {\bf (A)} or in case 
 {\bf (C)} the brane can only have the 
 first two behaviors. Case {\bf (E)} is 
 the only case with all four possibilities. 
 We also see that quite like the solutions
 obtained in \cite{Kutasov:2004dj}, here again 
 we find no solution describing the brane 
 in a stable orbit around the origin.

 Coupling to RR fields only happens when 
 $p'=p$ i.e. in case 2. The effects of 
 nonzero RR fields also appear in the 
 dependence of the solutions on the 
 value of $p$. In the approximate solutions
 which are obtained for $u\ll1$,
 this dependence vanishes. This is consistent
 with the fact that the effects of nonzero 
 RR fields become unimportant in the infrared 
 regime \cite{RR2}.
 As a result, the solutions we find in these 
 cases are quite similar to the ones found in
 \cite{Kutasov:2004dj} with the general behavior
 $r\sim{\frac{1}{\cosh{t}}}$ for small values 
 of $r$. We also found two solutions in 
 cases {\bf (C)} and {\bf (D)},
 where $j=0$, which are exact
 regardless of the initial conditions. 
 This was not possible in the background 
 (\ref{bac}).

 As was mentioned in \cite{Kutasov:2004dj}, 
 for purely radial motions of a D-brane 
 in the NS5-branes background one can,
 by a proper field redefinition, write the
 effective action as a DBI action for the 
 tachyon. In our problem, for example 
 in case 1 and for $L=0$, defining
 \be
 T=\sqrt{N}\;\ln{u}\;,
 \ee
 one can write the action as
 \be
 S_T=-\int\;d^{p'+1}\xi\;V_{p'}(T)\;
 \sqrt{1-\dot{T}^2}\;,
 \label{tacact}
 \ee
 where
 \be
 V_{p'}(T)=\tau_{p'}\;e^{\frac{T}{\sqrt{N}}}
 (1+e^{\frac{2T}{\sqrt{N}}})^{\frac{j}{4}}\;.
 \label{tacpot}
 \ee
 As $u\rightarrow0$, $T\rightarrow-\infty$ and
 as $u\rightarrow\infty$, $T\rightarrow\infty$.
 We now look at the 
 behavior of the potential in these two limits

 \bea
 {\frac{1}{\tau_{p'}}}V_{p'}(T)&\simeq&
 e^{\frac{T}{\sqrt{N}}}\;\;\;\;
 \;\;\;\;\;\;\;\;\;\;T\rightarrow-\infty\;.
 \eea
 As expected, in the limit $T\rightarrow-\infty$
 ($u\rightarrow0$), the situation is quite 
 similar to the one studied in \cite{Kutasov:2004dj} 
 and the potential  goes exponentially to 
 zero which is the late time behavior of the
 tachyon potential for unstable D-branes. 
 In the opposite limit, due to the nonzero RR 
 fields, we expect to see deviations from the
 long range gravitational attraction between 
 the D-brane and the fivebranes, $V(T)\simeq-1/T^2$, 
 observed in the background (\ref{bac}). 
 This is actually the case
 \bea
 {\frac{1}{
 \tau_{p'}}}V_{p'}(T)&\simeq&e^{\frac{T}{\sqrt{N}}}
 \;\;\;\;\;\;\;\;\;\;\;\;\;\;\;\;\;T\rightarrow
 \infty\;(j=0)\;,\cr &&\cr
 {\frac{1}{\tau_{p'}}}V_{p'}(T)&\simeq&1-{\frac12}
 e^{\frac{-2T}{\sqrt{N}}}
 \;\;\;\;\;\;\;\;T\rightarrow\infty\;(j=-2)\;.
 \label{larj}
 \eea

 For very small values of $u$, the potential 
 governing the dynamics of the D-brane is 
 that for a tachyonic field at late
 times. It is known that the pressure of a 
 tachyonic field drops exponentially to zero 
 at late times \cite{Sen:2002an}.
 In \cite{Kutasov:2004dj} it was shown that 
 this is also the case for a probe D-brane 
 near NS5-branes. For our problem also, as
 expected, this happens. To see this, we 
 just need to find the energy momentum
 tensor associated with (\ref{nac}) and read 
 the pressure
 \be
 P\sim-u(1+u^2)^{j/4}\sqrt{1+{\frac{N}{u^2}}
 \dot{u}^2}\;.
 \ee
 Plugging $u$ from (\ref{soljz}) in the above 
 expression results in the following behavior 
 for pressure at late times
 \be
 P\sim e^{-2t/\sqrt{N}}\;.
 \ee

 In the next section, we will ask if this 
 similarity persists when we use the worldvolume 
 theory to construct a cosmological model. 
 Motivated by this, we couple the worldvolume 
 theory of our probe D$_3$-brane to gravity 
 by introducing a four dimensional dynamical 
 background metric $g_{\mu\nu}$ which
 we choose to be a FRW solution of Einstein's 
 equations. We will then compare the resulting 
 models with the ones arising from tachyon driven 
 cosmologies. We will do this analysis both before
 and after the NS5-branes background is deformed 
 by RR fields.

 \section{Coupling to gravity}
 The worldvolume theory on a D$_5$-brane 
 moving in the NS5 branes background or 
 its deformation can be coupled to  gravity by
 truncating the space to a finite range of 
 values for the radial coordinate. The region 
 is then smoothly glued to a proper compact
 manifold to the outside space \cite{Kachru:2003sx}.
 As a result of this truncation, the effective 
 theory will include an Einstein-Hilbert term 
 in the action \cite{Verlinde:1999fy}.
 The resulting theory will describe a 
 six dimensional cosmology. If we insist 
 on having a four dimensional model, we 
 must start with the background where
 two of the six noncompact directions of the 
 NS5 branes are wrapped around a compact 
 manifold. 
 
 For example in \cite{aa} a D$_5$-brane has been
 considered in the presence of NS5-branes where both are wrapped
 around a 2-sphere and therefore the potential on the D-brane is
 that induced by Maldacena-Nunez background \cite{MN}. One could as well consider
 a D$_3$-brane in the MN background as a system which can couple to
 four dimensional gravity. We will discuss this scenario in section 3.3.
 
 Now let us see if and how the RR deformation of
 NS5-branes background that we considered in the previous
 sections can couple to four dimensional gravity. For this purpose
 we must note that, as stated in \cite{RR2}, one can
 find this background by considering M5-branes of 11 dimensional
 SUGRA in the presence of a $C$ field where the branes are smeared
 in a transverse direction. If one reduces this direction then the
 IIA NS5-branes background with an electric 3 form is obtained.
 The other solutions with different forms can then be found by
 T-duality in the directions where the magnetic forms are defined
 namely in the directions $p+1,\cdots,5$ of (\ref{nbc}). Upon taking the
 decoupling limit the background of our interest is obtained.

 Therefore it is implicit in this construction that the magnetic
 directions are considered on a torus. This can also be
 understood by the fact that the magnetic part of the metric
 shrinks by a factor of $1/ar$ for large values of $a$. As a result, one might
 suspect that a probe D$_3$-brane in the electric directions $0,\cdots,3$ can
 couple to four dimensional gravity\footnote{We thank M. Alishahiha for 
 useful comments on this point}
 . For other cases this
 reasoning clearly breaks down. A full account of this problem and
 the way the effective potential on the probe brane is modified,
 due to compactifying the two transverse directions of
 the D$_3$-brane, requires a more careful study which we postpone
 to a future work.

 In the following we take the first steps towards this problem and
 construct a four dimensional
 model by considering the effective action 
 on a probe D$_3$-brane moving in the NS5 or 
 its deformed background where the flat
 metric is replaced by a dynamical one. We 
 also assume that by a KKLT like 
 mechanism \cite{Kachru:2003aw} a nonzero 
 cosmological constant is included in the
 action. In fact we $\it{"assume"}$ that 
 there is a certain string
 compactification that results in such a model. 
 We also take all the moduli to be fixed.
 For the dynamical metric, we take the following form
 \be
 ds^2=-dt^2+a(t)^2(\frac{dr^2}{1-k\,r^2}
 +r^2\,d\Omega_2^2)
 \;\;\;\;,\;\;\;\; k=0,\pm1\;,
 \label{frw}
 \ee
 where $r$ is the radial coordinate inside 
 the brane.
 To make contact with the tachyon problem, by 
 proper field redefinition we will introduce 
 a tachyon  field $T$, in terms of which
 the action is brought into the general form
 \be
 S=\frac{1}{16\pi
 G}\int d^4 x\sqrt{-g}\,R-\int
 d^4x\sqrt{-g}\,\bigg(V(T)\sqrt{1-\dot{T}^2}
 +\Lambda\bigg)\;,
 \label{equ}
 \ee
 where $G\sim g_s^2$ is the four dimensional 
 gravitational coupling and
 $V\sim 1/g_s$. The potential $V(T)$ is 
 determined in each case and is
 to be compared with the potential 
 responsible for the decay of
 unstable D-branes \cite{garousi},\cite{effective}
 \be
 V_{tac}(T)=\frac{V_0}{\cosh{{\frac{T}{\sqrt{2}}}}}\;,
 \ee
 where we assume $V_0\gg\Lambda$. In each of 
 the following cases we will
 compare the results with the
 tachyon problem studied in \cite{Sen} whose 
 steps we follow
 closely. 
 For this purpose, we first write down the 
 equations of motion for $a(t)$ and $T(t)$ 
 arising from (\ref{equ})
 \be
 \frac{\ddot{a}}{a}=\frac{8\pi
 G}{3}\bigg[\Lambda+\frac{V(T)}{\sqrt{1-\dot{T}^2}}
 (1-{\frac32}\;\dot{T}^2)\bigg]\;,
 \label{eqa}
 \ee
 \be
 \ddot{T}=-(1-\dot{T}^2)\bigg[\frac{V'(T)}{V(T)}
 +3\;\dot{T}\;\frac{\dot{a}}{a}\bigg]\;.
 \label{eqt}
 \ee
 There is a constraint equation, known as 
 Friedman equation, which will read
 \be
 \bigg(\frac{\dot{a}}{a}\bigg)^2=-{\frac{k}
 {a^2}}+\frac{8\pi
 G}{3}\bigg[\frac{V(T)}{\sqrt{1-\dot{T}^2}}
 +\Lambda\bigg]\;.
 \label{cons}
 \ee
 In the above equation, $k/a^2$ is known 
 as the curvature term and
 $V(T)/\sqrt{1-\dot{T}^2}$ is the energy 
 density associated with
 the field $T$.
 Out of the four initial conditions, 
 $a_0,\dot{a}_0,T_0$ and
 $\dot{T}_0$ (evaluated at $t=0$), required 
 to solve the equations of motion,
 one, which we choose to be $a_0$, is obtained 
 in terms of the
 other three through (\ref{cons}).
 We can set $\dot{T}_0=0$ by choosing a proper 
 time origin and
 thus we are left with a two parameter family 
 of solutions which
 are labeled by $T_0$ and $\dot{a}_0$. Motivated 
 by the decay of
 an unstable D-brane which is a time reversal 
 symmetric process,
 we will only consider a one parameter subspace 
 of the possible
 solutions labeled by $\dot{a}_0=0$ for which
 we have the time reversal symmetry. But this 
 is not the most
 general solution of the equations of motion 
 and obtaining the full
 solutions is of course an interesting 
 generalization.

 The above initial conditions require that $k=1$. 
 for $k=-1$,
 $a_0$ is imaginary and for $k=0$, $a_0=0$ and 
 the model starts
 with singularity. Therefore for $k=1$ we have
 \be
 a_0=\bigg[\frac{8\pi G}{3}(V(T_0)+\Lambda)
 \bigg]^{-1/2}\;.
 \label{a_0}
 \ee

 \subsection{NS5 geometry}
 In this case as was mentioned 
 in \cite{Kutasov:2004dj}, $T(r)$ is 
 introduced by the relation
 \be
 {\frac{dT}{dr}}=\sqrt{h(r)}\;,
 \ee
 where $h(r)=1+{\frac{N}{r^2}}$\footnote{for 
 simplicity we will
 take $l_s=1$ hereafter.}. The tachyon 
 potential $V(T)$ is defined by
 \be
 V(T)=\frac{V_0}{\sqrt{h(r(T))}}\;,
 \ee
 where $V_0=\tau_3/ g_s$. As $r\rightarrow0$,
 $T\rightarrow-\infty$ and as $r\rightarrow
 \infty$,$T\rightarrow\infty$. In these limits 
 the potential behaves as
 \bea
 V(T)&\simeq&V_0e^{{\frac{T}{\sqrt{N}}}}
 \;\;\;\;\;\;\;\;\;\;\;\;\;\;\;\;\;\;\;\;
 \;T\rightarrow-\infty\;,\cr
 &&\cr
 V(T)&\simeq&V_0(1-{\frac{N}{2T^2}})
 \;\;\;\;\;\;\;\;\;\;\;\;T\rightarrow\infty\;.
 \eea
 This potentials has the same 
 behavior as $V_{tac}$ at late 
 times ($T\rightarrow-\infty$)
 but looks quite differently in the large $T$ 
 limit.

 Despite this difference, the 
 analysis of the solutions of
 (\ref{eqa}) and (\ref{eqt}) goes in
 close analogy with the discussion 
 presented in \cite{Sen}, having
 in mind that $|T_0|\ll1$ for 
 negative values of $T_0$ in the
 tachyon problem corresponds to 
 $T_0/\sqrt{N}\gg1$ in our problem.
 This is expected because in this 
 limit the potential $V(T)$ goes
 asymptotically to the value $V_0$ 
 which is the top of the
 potential. If it were that $V$ 
 could acquire this value, $T$
 would not evolve and the solution 
 for $a(t)$ would read
 \be
 a(t)=a_0\cosh{\bigg({\frac{t}{a_0}}\bigg)}\;.
 \label{acos}
 \ee
 But the potential can get at most 
 very close to $V_0$ for
 very large $T_0$ and (\ref{acos}) becomes an
 approximation. A quantity of interest 
 is $V'/V$ at $t=0$.
 This term that appears on the right 
 hand side of equation
 (\ref{eqt}), is the characteristic of 
 how flat the potential for rolling is.
 The smaller is
 $V'/V$, the better an approximation (\ref{acos}) 
 becomes at the beginning
 of the process. For large values of 
 $T$, $V'/V\sim1/T^3$.

 By the same reasoning of \cite{Sen}, as $T$ 
 evolves
 and $\dot{T}$ approaches its limiting value 
 of $-1$, which is
 the attractive fixed point of the 
 equations (for small enough
 $g_s$, the negative term $\dot{T}\dot{a}/a$ 
 is small compared to
 $V'/V$ and the right hand side of (\ref{eqt}) 
 remains negative), the relative
 magnitude of the cosmological constant 
 term $\Lambda$ and
 the curvature term at the point where the energy
 density and curvature become of the same 
 order (the cross-over
 point), determines
 whether the solution describes an open or 
 a closed universe. If
 $\Lambda$ is large compared to the curvature 
 at this point the
 expansion continues and if not it will 
 eventually stop. For larger values of $T_0$,
 the exponential like expansion of $a(t)$ 
 is a better approximation and
 the curvature term will be smaller at the 
 cross-over point.
 Therefore for a given $\Lambda$, there is 
 a critical value for
 the scalar field $T_{c1}$ such that for 
 $T_0>T_{c1}$ the universe is
 open.

 For sufficiently large negative values 
 of $T_0$, the
 initial value of the curvature term 
 becomes of the same order as
 the cosmological constant term. As $T$ 
 evolves, the curvature
 becomes smaller (although not very rapidly 
 because the potential
 is not flat enough and the near exponential 
 expansion is not a good approximation)
 and thus the right hand side of (\ref{cons})
 always remains positive and expansion 
 continues. Therefore there is a
 second critical value for the scalar 
 field $T_{c2}$ such that for
 $T_0<T_{c2}$ the universe is again 
 open and for the intermediate
 values $T_{c2}<T_0<T_{c1}$, the universe 
 is closed.
 By time reversal symmetry, the closed
 universe solution describes both a 
 big bang and a big crunch
 whereas the other one describes neither.

 \subsection{Deformed NS5 geometry}
 The background (\ref{nbc}) produces 
 effective theories on a probe D$_3$-brane 
 which were discussed in previous sections.
 We now couple these theories to gravity 
 for the two cases of section 2.
 
 $ $
 
 {\bf Case 1 :}
 In this case, as was mentioned in 
 section (2.5), the potential is

 \be
 V(T)=V_0\;e^{T/\sqrt{N}}
 (1+e^{2T/\sqrt{N}})^{j/4}\;,
 \ee
 where $V_0=\tau_3/\tilde{g}$ and $j=0,-2$
 \footnote{$j=-2$ only occurs in case 3 
 of section 2 with (p,s,q) given by (1,0,4)
 or (3,1,5).}.
 For $T\rightarrow-\infty$ the potential 
 goes as $e^{T/\sqrt{N}}$
 regardless of $j$ and hence our 
 discussion of the previous section
 remains unchanged in this limit. However, for 
 large values of
 $T$ and as seen in (\ref{larj}), depending on 
 the value of $j$,
 the potential takes two different forms
 both of which look differently from what we 
 saw in the previous
 section.

 For $j=0$ we see that $V'/V=1$. So even for 
 large values of $T_0$,
 the near exponential growth of $a(t)$ is not 
 a good approximation and
 the cross-over happens very rapidly. By 
 choosing larger values for $T_0$,
 $a_0$ will be smaller and therefore the 
 curvature term decreases with a
 bigger slope immediately after the evolution 
 starts.
 Hence we expect $T_{c1}$ to become large. As a 
 result, the
 range of values of $T_0$ for which the solution 
 becomes singular
 is much wider in this case. For $j=-2$ and for 
 large values of
 $T$, $V'/V\sim e^{-2T}$.
 
 $ $
 
 {\bf Case2 :}
 In this case the effective action 
 contains an additional term
 coming from RR coupling and the 
 action (\ref{equ}) and the
 equations resulting from that are 
 modified. The action is
 \be
 S=\frac{1}{16\pi G}\int d^4 x\sqrt{-g}\,
 R-\int d^4x
 \sqrt{-g}\bigg(V_1(T)\sqrt{1-\dot{T}^2}
 -V_2(T)+\Lambda\bigg)\;,
 \ee
 where
 \bea
 V_1(T)&=&V_0\;e^{T/\sqrt{N}}(1+e^{2T/
 \sqrt{N}})^{1/2}\;, \cr &&\cr
 V_2(T)&=&V_0\;e^{2T/\sqrt{N}}\;,
 \eea
 and $V_0=\tau_3/\tilde{g}$. The 
 equations of motion for $a(t)$ and $T(t)$ are
 \be
 \frac{\ddot{a}}{a}=\frac{8\pi
 G}{3}\bigg[\Lambda+\frac{V_1(T)}
 {\sqrt{1-\dot{T}^2}}(1-{\frac32}\;
 \dot{T}^2)-V_2(T)\bigg]\;,
 \label{eqac2}
 \ee
 \be
 \ddot{T}=-(1-\dot{T}^2)\bigg[
 \frac{V_1'(T)}{V_1(T)}+3\;
 \dot{T}\;\frac{\dot{a}}{a}-
 \sqrt{1-\dot{T}^2}\frac{V_2'(T)}{V_1(T)}
 \bigg]\;.
 \label{eqtc2}
 \ee
 The Friedman equation is
 \be
 \bigg(\frac{\dot{a}}{a}\bigg)^2=
 -{\frac{k}{a^2}}+\frac{8\pi
 G}{3}\bigg[\frac{V_1(T)}{\sqrt{1-
 \dot{T}^2}}-V_2(T)+\Lambda\bigg]\;.
 \label{consc2}
 \ee
 Here, the combination $(V'_1-V'_2)/V_1$ 
 determines  how slowly $T$
 evolves at the beginning of its rolling. For 
 large values of $T$
 it goes as $e^{-4T/\sqrt{N}}$ and thus in 
 this limit the potential for
 rolling is flat to a good approximation. 
 Furthermore, as $T$ evolves, the combination 
 $V_1/\sqrt{1-\dot{T}^2}-V_2$ 
 as compared to the
 energy density of the above cases remains
 almost constant for a longer period of time.
 On the whole, in comparison with the 
 previous cases, on the one hand $a(t)$
 grows almost exponentially to a better 
 approximation and on the other, the
 cross-over is postponed to a later time 
 resulting in a smaller value for
 $T_{c1}$.

 We also note that for large negative 
 values of $T$, the initial
 curvature and cosmological constant 
 terms become of the same
 order for smaller magnitudes of $T_0$ 
 in comparison with the
 other cases thus resulting in a larger 
 value for $T_{c2}$. Therefore
 $T_0$ can take a narrower range of 
 values to make a singular
 solution.
 
 \subsection{MN Geometry}
 In this subsection we study a probe 
 D$_3$-brane in the MN background (see Appendix A) which is
 parallel to the flat directions of the metric. 
  
 For a purely radial motion of the probe it is easy to verify that the 
 effective action in terms of the tachyon field $T\equiv\sqrt{N}r$ is 
 written in the form of (\ref{tacact}) with $p'=3$ and
 \be
 \frac{1}{\tau_3}V(T)=\frac{\sinh^{\frac{1}{2}}\frac{2T}{\sqrt{N}}}
 {\bigg(4\frac{T}{\sqrt{N}}\coth\frac{2T}{\sqrt{N}}-
 4\frac{T^2}{N\sinh^2\frac{2T}{\sqrt{N}}}-1\bigg)^{\frac{1}{4}}}\;.
 \ee
 Note however that the tachyon field varies between zero and infinity this time.
 We then obtain the following asymptotic behaviour for the potential 
 \bea
 \frac{1}{\tau_3}V(T)&\simeq&1+\frac{4}{9N}T^2\,\,\,\,\;\;\;\;\;\;\;\;
 \;\,\,\,\,T\sim 0\;, \cr &&\cr
 \frac{1}{\tau_3}V(T)&\simeq&\frac{1}{2}\bigg(\frac{T}{\sqrt{N}}
 \bigg)^{-\frac{1}{4}}e^{\frac{T}{\sqrt{N}}}\,\,\,\,\,\,\,\,T\rightarrow +\infty\;.
 \eea
 For large values of $T$ we see that the behavior is more or less similar to
 some of the results obtained so far namely case 1 with $j=0$.
 In this region one obtains $V'/V\sim 1$ and, as stated before, exponential
 expansion is not a good approximation.
 For small values of $T$, which would correspond to $T\rightarrow-\infty$ of the 
 previous cases, although the potential does not become zero its derivative does
 which is similar to the late time behaviour of tachyon potential.
 In fact $V'/V\sim T$ for small $T$
 and therefore $a(t)$ grows near exponentially with a very good approximation
 in this region (see (\ref{eqa}) and (\ref{eqt})). This causes $T_{c_2}$ to become
 larger in comparison with the previous cases and the singular window to become narrow.  
 A numerical analysis of the case confirms these arguments completely.

 \subsection{Comments on the solutions}
 The solutions we found for the NS5 geometry and its deformations 
 show qualitatively the same behavior. For 
 all the cases two critical
 values, $T_{c1}$ and $T_{c2}$ are 
 defined such that for
 $T_0>T_{c1}$ and $T_0<T_{c2}$ the 
 solutions are nonsingular
 whereas for $T_{c2}<T_0<T_{c1}$ we 
 will have singular solutions.
 Having in mind that we are dealing 
 only with time reversal 
 symmetric solutions, we note that 
 the solutions with singularity 
 describe a big bang in the past 
 and a big crunch in the future and for
 the nonsingular solutions neither 
 of the two exists.
 We also note that we should necessarily 
 have a nonzero
 cosmological constant in order to allow 
 for nonsingular
 solutions (see \cite{alex} for a 
 supergravity  analysis of a similar 
 model where all the solutions are 
 singular in 4D).   
 While the potentials for the different cases 
 look almost the same way as
 $T\rightarrow-\infty$, they behave rather 
 differently in the
 opposite limit. As a result, $T_{c2}$ has 
 almost the same value
 for all the cases but the values 
 for $T_{c1}$ depend rather
 considerably on the case. Hence, the 
 RR deformation affects the
 range of values $T_0$ can take in 
 order to result in a singular
 solution.

 For the values of $\frac{8\pi G\Lambda}{3}
 =.001$ and $\frac{8\pi G\tau_3}{3g_s}=.05$
 we have found by numerical analysis 
 that $(T_{c2}/\sqrt{N},T_{c1}/\sqrt{N})$ for
 the above cases of NS5 geometry, $j=-2,j=0$ 
 and case $2$ are
 respectively $(-2.7,1.3),(-3.0,0.9),(-3.0,2.6)$ 
 and $(-2.8,-0.1)$. We have also 
 plotted $a(t)$ for $j=0$ and for the
 initial values of $T_0=-4,1$ and $3$ in 
 figures 1 to 3 respectively. In figure 4, 
 $T$ as a function of time is plotted
 for the three different initial values. For the MN geometry, we 
 take a smaller values for the above constants in our numerical calculation
 in order to make the singular window visible (see figure 5).

 As an important point one 
 should note that unlike the potential for 
 rolling tachyon $V\sim1/\cosh{T}$, 
 the potentials we have been dealing with
 meet the criteria for slow roll inflation 
 in the  large $T$ limit.
 In this limit, where $\dot{T}$ and $\ddot{T}$ 
 are small and for $V_0\gg\Lambda$, the 
 slow roll parameters $\epsilon$ and $\eta$
 are written as
 \be
 \epsilon=\frac{m_p^2}{2}\frac{V'}{V^3}
 \;,\;\;\;\;\;\;\;\;
 \eta+\epsilon=m_p^2\frac{V''}{V^2}\;,
 \ee
 where $m_p^2=1/\sqrt{8\pi G}$. It is 
 easy to see that for sufficiently 
 large values of $T$, the above
 parameters are small for the cases of 
 NS5 geometry, $j=-2$ and $j=0$.
 For case $2$ however, the parameters are 
 found as
 \be
 \epsilon=\frac{m_p^2}{2}\frac{V'^2}
 {V_1V^2}\;,\;\;\;\;\;\;\;
 \eta=m_p^2\frac{V''}{VV_1}
 -\epsilon\frac{VV_1'}{V'V_1}\;,
 \ee
 where $V=V_1-V_2$. Here again the 
 slow roll conditions are satisfied 
 for large $T$.
  
 It is important to note that all 
 the above solutions and results
 are only reliable if the string 
 coupling remains small during the
 rolling. Therefore, for a given 
 $g_s$ or $\tilde{g}$, not all the
 initial values of $T_0$ are allowed. Even 
 if this requirement is met
 in the beginning of the rolling, all the 
 solutions will sooner or
 later violate the condition. We can at 
 best postpone this
 violation to late times like what we 
 did in section 2. For this
 purpose, we require the value of the 
 coupling to be extremely
 small at $t=0$. For the RR deformed 
 cases this will read
 \be
 \tilde{g}\frac{(1+e^{2T_0/\sqrt{N}})^
 {\frac{p-1}{4}}}{e^{T_0/\sqrt{N}}}\ll1\;.
 \ee
 For large negative values of $T_0$ the 
 above condition, including
 the NS5 geometry case, becomes
 \be
 g\ll e^{T_0/\sqrt{N}}\ll1\;,
 \ee
 where $g$ is $g_s$ for the NS5 geometry 
 and $\tilde{g}$ for the
 deformed one. This relation is the analogue of the
 condition $g\ll\lambda\ll1$ of section 2.

 \section*{Acknowledgments}
 We would like to thank Farhad Ardalan, 
 Alex Buchel and Mohammad
 Reza Garousi for very  useful comments 
 and Hossein 
 Hakimi Pajouh for helping with the 
 numerical calculations. We would
 especially like to thank Mohsen 
 Alishahiha for valuable
 discussions via email.
 
 \appendix
 \section{Maldacena-Nunez background}
 
 The MN background for the metric,  the NS three form and the dilaton is given by 
 \bea
 &&ds^2=dx_4^2+N \left(d\rho^2+e^{2g(\rho)} d\Omega^2_2+\frac{1}
 {4}\sum_a(\omega_a-A_a)^2\right)\,,\cr &&\cr 
 &&H_3=N\left[-\frac{1}{ 4}(\omega_1-A_1)\wedge(\omega_2-A_2)\wedge(\omega_3-A_3)
 +\frac{1}{ 4} \sum_a F_a\wedge(\omega_a-A_a)\right] \,,\cr &&\cr
 &&e^{2(\phi-\phi_0)}=\frac{2e^{g(\rho)}}{\sinh 2\rho}\,,\,\,\,\,
 e^{2g(\rho)}=\rho\coth2\rho-\frac{\rho^2}{\sinh^2(2\rho)}-\frac{1}{4}
 \eea
 where $\Omega_2^2$ is a two sphere parametrized by 
 $(\tilde{\theta},\tilde{\phi})$ where
 $NS5$-branes are wrapped and $\omega_a$ are the $SU(2)$ 
 left-invariant one forms on the
 three sphere parametrized by $(\theta,\phi,\psi)$,
 \bea
 \omega_1&=&\cos\phi\ d\theta+\sin\phi\sin\theta d\psi\cr &&\cr
 \omega_2&=&-\sin\phi\ d\theta+\cos\phi\sin\theta d\psi\cr &&\cr
 \omega_3&=&d\phi\ +\cos\theta d\psi \,.
 \eea
 and $A_a$ are the $SU(2)_R$ gauge fields on the two sphere
 \bea
 A_1&=&a(\rho)\ d\tilde{\theta}\,\,\,\,,\,\,\,\,A_2=a(\rho) 
 \sin\tilde{\theta}\  d\tilde{\phi}\,\,\,\,,\,\,\,\,
 A_3=\cos\tilde{\theta}\ d\tilde{\phi}\,, \cr &&\cr
 F_a&=&dA_a+\frac{1}{2}\epsilon_{abc} A_b\wedge  
 A_c\,\,\,\,,\,\,\,\,a(\rho)=\frac{2\rho}{\sinh2\rho}\,,
 \eea
 \begin{figure}[f1]
 \begin{center}
 \epsfig{file=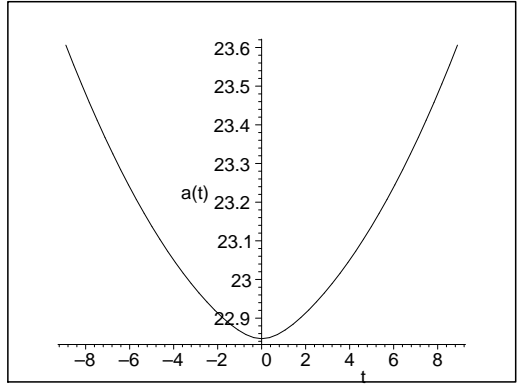,width=0.45\textwidth}
 \caption{$\frac{8\pi G\Lambda}{3}=.001$ 
 and $\frac{8\pi G\tau_3}{3g_s}=.05$.
 For $T_0=-4$ the universe is open.}
 \label{fig1}
 \end{center}
 \end{figure}

 \begin{figure}[f2]
 \begin{center}
 \epsfig{file=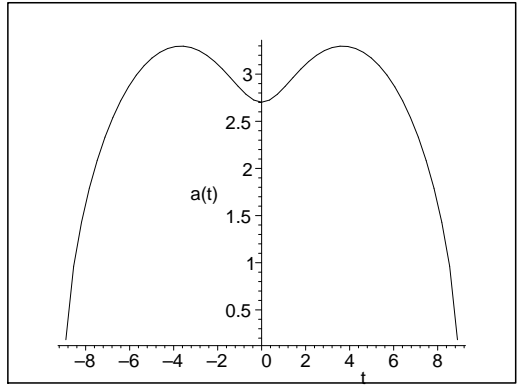,width=0.45\textwidth}
 \caption{ $T_0=1$ results in a big 
 bang and a big crunch.}
 \label{fig2}
 \end{center}
 \end{figure}

 \begin{figure}[f3]
 \begin{center}
 \epsfig{file=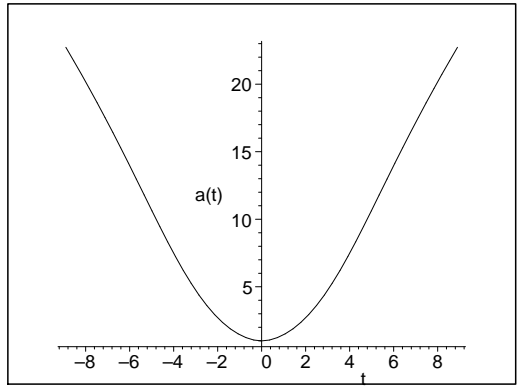,width=0.45\textwidth}
 \caption{ $T_0=3$ results in an open 
 universe.}
 \label{fig3}
 \end{center}
 \end{figure}

 \begin{figure}[f4]
 \begin{center}
 \epsfig{file=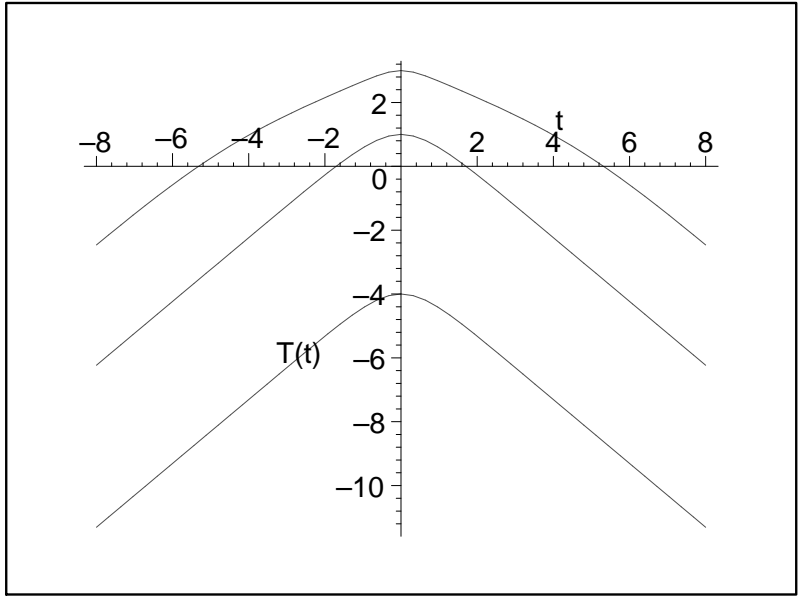,width=0.7\textwidth}
 \caption{$T(t)$ for different initial 
 conditions.}
 \label{fig4}
 \end{center}
 \end{figure}

 \begin{figure}[f5]
 \begin{center}
 \epsfig{file=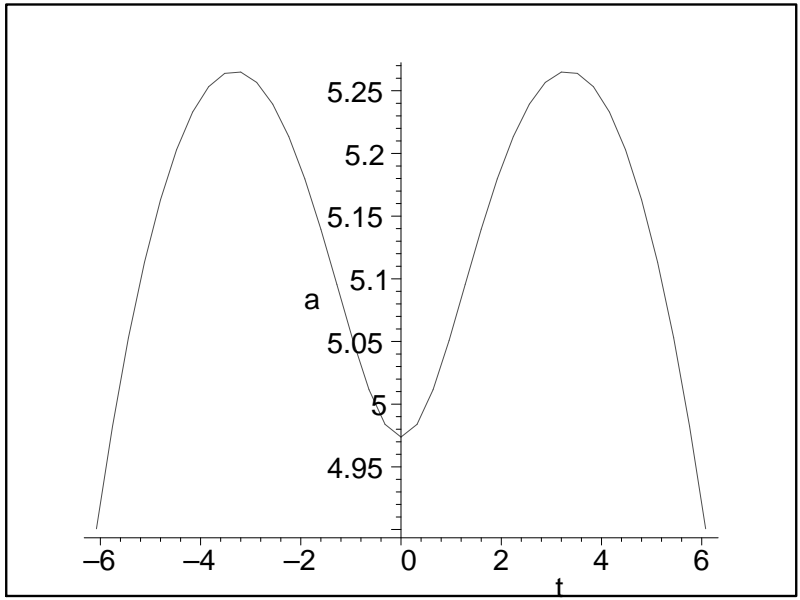,width=0.7\textwidth}
 \caption{$a(t)$ for MN background with $\frac{8\pi G\Lambda}{3}=.00001$ 
 and $\frac{8\pi G\tau_3}{3g_s}=.0005$.
 For $T_0=5.5$ the universe is closed.}
 \label{fig5}
 \end{center}
 \end{figure}

 \end{document}